\documentclass[12pt,preprint]{aastex}

\def\beq{\begin{equation}}
\def\eeq{\end{equation}}
\def\beqar{\begin{eqnarray}}
\def\eeqar{\end{eqnarray}}

\def\la{\mathrel{\mathpalette\fun <}}
\def\ga{\mathrel{\mathpalette\fun >}}
\def\fun#1#2{\lower3.6pt\vbox{\baselineskip0pt\lineskip.9pt
  \ialign{$\mathsurround=0pt#1\hfil##\hfil$\crcr#2\crcr\sim\crcr}}}

\begin{document}

\title{Neutrino Constrains to the Diffuse Gamma-Ray Emission from Accretion Shocks}

\author{ A. Dobard\v zi\'c}
\affil{Department of Astronomy, Faculty of Mathematics, University of Belgrade, Studentski trg 16, 11000 Belgrade, Serbia}
\email{aleksandra@matf.bg.ac.rs}

\and

\author{T. Prodanovi\'c}
\affil{Department of Physics, University of Novi Sad, Trg Dositeja Obradovi\'ca 4, 21000 Novi Sad, Serbia}
\email{prodanvc@df.uns.ac.rs}

\begin{abstract}

Accretion of gas during the large scale structure formation has been thought to give rise to shocks that can
accelerate cosmic rays. This process then results in an isotropic extragalactic gamma-ray emission contributing to the
extragalactic gamma-ray background observed by the \emph{Fermi}-LAT. Unfortunately this emission has been difficult to constrain and thus presents an uncertain foreground to any attempts to extract potential dark matter signal.
Recently, IceCube has detected high-energy isotropic neutrino flux which could be of an extragalactic origin.
In general, neutrinos can be linked to gamma rays since cosmic-ray interactions produce neutral and charged
pions where neutral pions decay into gamma rays, while charged pions
decay to give neutrinos. By assuming that isotropic high-energy IceCube
neutrinos are entirely produced by cosmic rays accelerated in accretion shocks during the process of structure formation, we obtain the strongest
constraint to the gamma-ray emission from large scale structure formation (strong) shocks and find that they can make at best $\sim 20\%$
 of the extragalactic gamma-ray background, corresponding to neutrino flux with spectral index $\alpha_\nu = 2$, or $\sim 10\%$ for spectral index $\alpha_\nu = 2.46$.
Since typical objects where cosmic rays are accelerated in accretion shocks are galaxy clusters, observed high-energy neutrino fluxes can then be used
to determine the gamma-ray emission of a dominant cluster type and constrain acceleration efficiency, and thus probe the process of large scale structure formation.
\end{abstract}


\section{INTRODUCTION}
\label{sec:intro}

Interactions of cosmic rays accelerated in various astrophysical environments  result in production of both charged and neutral
pions through $pp$ collisions.
Although charged pions decay producing neutrinos
$\pi^{+}\rightarrow \nu_\mu \overline{\nu}_\mu \nu_e e^{+}$, $\pi^{-}\rightarrow \overline{\nu}_\mu \nu_\mu
\overline{\nu}_e e^{-}$~\citep{MSS78,ST79,MWW90}, while neutral pions decay into gamma rays $\pi^0\rightarrow \gamma\gamma$
 ~\citep{STE70,STE71}, both originate from the same cosmic-ray source and thus observations of resulting neutrinos and gamma rays can
 be linked to give us a more detailed picture of  the source cosmic rays.
Neutrinos are especially important tracers of acceleration processes, since they travel
 long distances without absorption or magnetic deflection. Recent high-energy neutrino detection reported by the IceCube collaboration
 includes $37$ events with energies ranging from $60\mathrm{TeV}-3\mathrm{PeV}$~\citep{AAA2014}. These high-energy events are
 best fitted by hard spectra $E_\nu^{-2}$ and the best fit single flavor ($\nu_i+\bar{\nu_i}$) neutrino flux (where $i=e,\mu,\tau$) in this energy range is
 $E_{\nu_i}^2 I_{\nu_i}(E_{\nu_i})=0.95\pm0.3\times10^{-8}\,\mathrm{GeV}\mathrm{cm}^{-2}\mathrm{s}^{-1}
 \mathrm{sr}^{-1}$~\citep{AAA2014}. These high-energy neutrino events were detected isotropically which suggest that their origin is either from common isotropically distributed sources or diffuse sources~\citep{AAA2014}. Sources of these neutrinos are still unknown but many have already been proposed such
as jets and cores of active galactic nuclei ~\citep{SDS91,AHST2008}, gamma-ray burst ~\citep{WB97,AHST2008,MI13}, starburst galaxies
 ~\citep{HWF13,MAL13,CW14,LWI14} and galaxy clusters \citep{ZTG14}.

Most recent all-sky gamma-ray observations have been performed by the \emph{Fermi}-LAT, which has
observed the diffuse gamma-ray sky in the energy range $0.1-820\,\mathrm{GeV}$~\citep{AC14}.
Besides the Galactic gamma-ray emission, a diffuse extragalactic gamma-ray background (EGRB) emission was also detected by \emph{Fermi}~\citep{AC14}.
Objects like unresolved blazars~\citep{SS96,DE07,NT07,IT09,SPA12}, high-latitude contamination by pulsar radiation~\citep{FL10},
dark matter annihilation~\citep{SC10,ABDO10}, secondary gamma-ray cascades~\citep{MBT12,II12}, unresolved star-forming
galaxies~\citep{ST76,PF02,PF06,MTK11}
were considered, and although some combinations of these components could explain the observed EGRB, large uncertainties are still present,
and thus the presence of additional component(s) has not yet been ruled out.

Here we investigate the case of cosmic rays expected to be accelerated in accretion shocks during the large-scale structure formation -- structure formation cosmic
rays (SFCRs; Loeb \& Waxman 2000;
Furlanetto \& Loeb 2004; Miniati et al. 2000). Though still hypothetical, this cosmic-ray population is expected to be present
at accretion and merger shocks, especially in connection to accretion shocks around clusters. This should result in gamma-ray
emission from hadronic~\citep{KW09,PP10} processes that dominate, and subdominant primary and secondary Inverse Compton leptonic emission ~\citep{PP10,LW00,FL04}, as well as radio emission dominated by its leptonic
component~\citep{EBP98,KKW09}. In this work we will focus on the hadronic emission component.
The collective emission from all unresolved clusters would also contribute to the extragalactic emission background, but as the
evolution of the sources is unknown and no sources have been detected the limits are
weak \citep{MI03,PF04,PF05,KU05}.
Signatures of SFCRs were expected to be detected by the \emph{Fermi}-LAT observations of
galaxy clusters, however only flux upper limits were
placed so far~\citep{AC10,AC14b}. In order to constrain their contribution to the EGRB, in~\cite{DP14} we have
constructed a
model of gamma-ray emission from large-scale accretion shocks around clusters implementing for the first time a source evolution
and normalizing
using \emph{Fermi}-LAT cluster observation limits. Due to large uncertainty in normalization of our models, the resulting limits were very weak
and  dependent on the choice of typical source i.e. a typical cluster size that dominates emission.
On the other hand, if a strong limit to SFCR contribution to the EGRB could be placed, our model could be very constraining for the
typical cluster-class that dominates the SFCR emission and also serve to probe different models of accretion shock evolution.

A significant flux of isotropic high-energy neutrinos has recently been detected by the IceCube.
The observed IceCube neutrinos cover a energy range $60\,\mathrm{TeV}-3\mathrm{PeV}$~\citep{AAA2014} and detected flux is best fitted with the hard spectrum $E_\nu^{-2}$. Since cosmic-ray interactions also result in neutrino production through charged pions, any extragalactic cosmic-ray population would produce extragalactic neutrino background as well as the diffuse gamma-ray background component. In the case of detected neutrino flux with spectral index $\alpha_\nu=2$ corresponding to cosmic-rays accelerated in strong shocks, ~\cite{CW14} calculated that these sub-PeV and PeV neutrinos should be accompanied by gamma-ray flux of $E_\gamma^2 I_\gamma(E_\gamma)=2\times10^{-8}\,\mathrm{GeV}\mathrm{cm}^{-2}\mathrm{s}^{-1}\mathrm{sr}^{-1}$. Such hard spectrum is consistent with emission expected from galactic cosmic-rays in starburst galaxies. This possibility was explored in \cite{TAM14}, and though
it can provide an explanation to the observed neutrino flux consistent with the observed EGRB, there remain issues regarding acceleration of protons
to such high energies and their confinement. On the other hand, accretion shocks that arise during the growth of structures cover a distribution of strengths~\citep{MI00} and SFCR population accelerated in them would also be consistent with such hard spectral index as follows from neutrino observations. Moreover, due to their large scale, accretion shocks would be more likely sites to accelerate and confine such energetic cosmic rays \citep{norman}.

Possibility of clusters as a source of observed high-energy neutrino flux has been also explored in several different works.
For example, \cite{MAL13} have considered the contribution of clusters, active galactic nuclei and star-forming galaxies is galaxy clusters  as sources contributing to neutrino background observed by the IceCube. Without including source evolution, they found that contributing galaxy clusters should have spectra with hard spectral index $\alpha\la 2.1-2.2$ contributing $\ga 30\%-40\%$ to the EGRB.
\cite{ZTG14} have modeled gamma-ray and neutrino backgrounds for a range of radio-loud galaxy cluster types. In their work \cite{ZTG14} have used evolving, numerically modeled cluster mass function from ~\cite{TKK08}, with cluster gamma-ray luminosity determined from: a) a phenomenological mass-gamma-ray-luminosity relation that was calibrated by radio observations of clusters with gamma-ray luminosity scaled of radio luminosity; b) a more sophisticated intracluster medium and cosmic-ray distributions.
\cite{ZTG14} have found that for the case of strong shock with $\alpha=2$, if all clusters are radio-loud they can account for at best $20\%$ of the observed neutrino background, or $10\%$ if only $30\%$ are radio loud, corresponding to only a few percent of the gamma-ray background. For softer spectra these numbers are even lower. In a more sophisticated model with percentage of radio-loud clusters ranging from $10-40\%$ where gamma-ray fluxes of radio-quiet ones were taken to be an order of magnitude lower than the fluxes of radio-loud ones, \cite{ZTG14} have found that SFCRs contribute $<1\%$ to neutrino and gamma-ray backgrounds.

Since SFCR population is yet to be detected and large uncertainties remain, in this work we will place constraints on the SFCR contribution to the gamma-ray background in a more model independent way. We use high-energy neutrinos recently detected by the Ice Cube as an upper limit to SFCR-made neutrino background flux from which we determine the corresponding gamma-ray background emission using the approach analyzed in \cite{DP14}. In \cite{DP14} we have constructed a model of gamma-ray background expected from cosmic rays accelerated in accretion shocks with evolution of the sources based on \cite{PAFI06} model of accretion shocks where evolution of the power processed by a distribution of accretion shocks was analyzed.
 Furthermore, gamma-ray background limit obtained this way can then be used to learn about the accretions shock emission and types of objects that it originates from.

\section{FORMALISM AND RESULTS}
\label{sec:form}

Hadronic interactions of cosmic-ray protons with ambient gas ($pp$ interactions) produce neutral and charged pions, which decay and produce gamma rays and neutrinos  $\pi^0\rightarrow \gamma\gamma$, $\pi^{+}\rightarrow \nu_\mu \overline{\nu}_\mu \nu_e e^{+}$, $\pi^{-}\rightarrow \overline{\nu}_\mu \nu_\mu
\overline{\nu}_e e^{-}$. In these reactions each gamma-ray takes half of the pion energy, while each of the three neutrinos from charged pion decay carries
about one quarter of the pion's energy (hence $E_\gamma=2E_{\nu_i}$). The relative flux of neutrinos and gamma rays depends on the relative number of produced charged and neutral pions, which is in the case of the $pp$ interactions $N_{\pi^\pm}/N_{\pi^0}\simeq2$~\citep{KK06,KA06}. After oscillations, the relative flavor ratio of neutrinos, as well as  anti-neutrinos, is $\nu_e:\nu_\mu:\nu_\tau=1:1:1$~\citep{KK06,KA06}. At a given energy $E$ this results in a simple connection~\citep{AM14,TAM14,CW14} between differential fluxes of all neutrino flavors and gamma rays $E^2I_\gamma(E) \simeq \frac{3E^2I_{\nu_i}(E)}{6/2^{\alpha}}$~\citep{AM14}, which in the case of strong shocks $\alpha=2$ becomes:
\begin{equation}
\label{gammanu}
E^2I_\gamma(E) \simeq 2E^2I_{\nu_i}(E)=\frac{2}{3}E^2I_{\nu}(E)\,,
\end{equation}
where $I_{\nu_i}$, $I_{\nu}$ and $I_\gamma$ are differential fluxes $I (E) \propto E^{-\alpha}$  of single flavor neutrinos, all-flavor neutrinos and gamma rays respectively. So, if we include all neutrino flavors and neglect any absorption of gamma rays and neutrinos during their propagation, we expect the differential gamma-ray flux to be around twice as high as that of a single flavor neutrinos for the case of strong shocks with  spectral index $\alpha=2$, and for the case of weaker shocks slightly higher even.

Using the best fit single flavor high-energy neutrino flux and spectrum as measured by the IceCube ~\citep{AAA2014} we can use the relation (\ref{gammanu}), which links gamma-ray flux and all-flavor neutrino flux, to find the corresponding gamma-ray flux. In the case of cosmic rays accelerated in strong-shocks with spectral index $\alpha={2} $ we get that the gamma-ray flux at some energy that corresponds to combined flux of all neutrino flavors at that same energy is $E^2 I_\gamma(E)=1.9\times10^{-8}\,\mathrm{GeV}\mathrm{cm}^{-2}\mathrm{s}^{-1}\mathrm{sr}^{-1}$. Assuming that cosmic rays in question are in fact SFCRs, i.e. that all of the high-energy IceCube neutrinos came from interactions of cosmological cosmic rays, we use the corresponding gamma-ray flux at TeV energies to determine the gamma-ray background emission resulting from SFCRs. This directly fixes the normalization of our model of SFCR gamma-ray emission from accretion shocks derived in~\cite{DP14}.
Models of SFCR gamma-ray emission from accretion shocks presented in ~\cite{DP14} depend on the choice of the normalizing cluster. This should be a typical cluster with respect to its SFCR gamma-ray emission i.e. a cluster class which process most of the accretion shock power (when averaged over the cosmic history) and are thus dominant sources of SFCRs. Therefore, when neutrino emission is used to fix the normalization, the resulting gamma-ray flux also fixes the SFCR gamma-ray flux from a typical cluster of galaxies.

If large-scale structure-formation shocks that process most of the gas are weak shocks, rather than strong as we have assumed so far,
then the spectral index of the SFCRs will be larger than that of observed high-energy  neutrinos $\alpha > 2$. Even then
we can use detected neutrino flux as the upper most limit that must not be overshot, and constrain maximal allowed SFCR-made
neutrino and gamma-ray flux for any given weak shock spectrum. In that case, however, cosmic rays accelerated in such shocks
would not be able to account for observed neutrino flux entirely due to difference in spectral indices, and additional sources would be needed. Nevertheless, we find that
in the case of weak shocks, even though all of the SFCR neutrino fluxes were adopted to be consistent with the IceCube data, for spectral
indices $\alpha\ga2.3$ all of the accompanying gamma-ray background fluxes violate the \emph{Fermi} data. Thus, for the case of SFCRs
accelerated in weak shocks with $\alpha\ga2.3$, \emph{Fermi} gamma-ray observations are more constraining than the detected IceCube
neutrino flux. In the case of sources with spectral index $\alpha \approx 2.3$ both gamma-ray and neutrino
background fluxes can be used in concert to place strongest constraint on  such SFCR contribution. For any
other spectral index either gamma-ray or neutrino background will be more constraining. However, if more then
one source besides standard known sources is considered, then again both gamma-ray and neutrino backgrounds
can be used together to limit their joint contribution on different ends - gamma-rays to constrain sources
with steep spectral indices and neutrinos to constrain sources with hard spectral indices.

Very recently an update to the diffuse neutrino flux detected by IceCube collaboration was reported for the energy range $25\,\mathrm{TeV}-1.4\,\mathrm{PeV}$~\citep{AAA15}, with the best fit single flavor neutrino spectral index of $\alpha_\nu=2.46$ and flux $I_{\nu_{i}}(E_{\nu_{i}})=2.6^{+0.4}_{-0.3}\times10^{-18}(E_{\nu_{i}}/10^5\,\mathrm{GeV})^{-2.46\pm0.12}$ $\mathrm{GeV}^{-1}\mathrm{cm}^{-2}\mathrm{s}^{-1}\mathrm{sr}^{-1}$. We complete our analysis for both reported neutrino fluxes in parallel.

Assuming that some or the entire detected high-energy neutrino flux originates from cosmic rays accelerated in structure-formation accretion shocks, we find the corresponding gamma-ray flux and determine its contribution to the EGRB for different assumed cosmic-ray spectra and plot these results on Figure 1.
Panels on the left show SFCR curves after normalization to the $\alpha_\nu=2.0$ neutrino spectrum from~\cite{AAA2014}, and panels on the right use normalization to latest $\alpha_\nu=2.46$ neutrino spectrum from~\cite{AAA15}. Top panels represent SFCR curves (dash dotted line) derived for the case of strong shocks i.e. for cosmic-ray spectral index $\alpha=2.0$ which matches that of neutrinos from~\cite{AAA2014}. Normalization to $\alpha_\nu=2.0$ neutrino spectrum from~\cite{AAA2014} yields an upper limit to the SFCR contribution of $\approx 46\%$ of the EGRB observed by the \emph{Fermi}, which is around the contribution of some of the major components like unresolved star-forming galaxies contributing $\la50\%$ (Fields et al. 2010; dashed line), or blazars contributing $\approx16\%$ (Abdo et al. 2010c; dotted line). If, for this same SFCR model, the attenuation of highest energy gamma-ray photons by extragalactic background light ~\cite[EBL;][]{GSP12} is included (thin solid line, and thin red solid line in the online journal), we find that this reduces the SFCR contribution to $\la 18\%$ of the observed EGRB. The sum of all components is shown with thick solid line (thick blue solid line in online journal).
Using the latest neutrino spectrum with $\alpha_\nu=2.46$~\citep{AAA15} as the upper limit constraint, we find that for SFCR accelerated in strong shocks i.e. with spectral index $\alpha=2.0$ their contribution to the EGRB can be $\approx29\%$, or $\la 12\%$ after the EBL attenuation was included. However, SFCRs with such spectrum cannot explain the entire detected high-energy neutrino flux and additional sources would be needed. If, on the other hand, SFCRs are assumed to have softer spectra with indices $\alpha>2.0$, their resulting fluxes would quickly violate the \emph{Fermi} EGRB observations. Middle panels of Figure 1 show SFCR gamma-ray fluxes for source spectra $\alpha=2.3$ while bottom panels correspond to $\alpha=2.6$. Here we point out that using the latest neutrino fluxes with spectrum $\alpha_\nu=2.46$~\citep{AAA15} is more constraining in the sense that it does not allow for this entire neutrino flux to be of the SFCR origin because the accompanying gamma-ray flux quickly violates the observed EGRB as we can see on bottom-two panels on the right side of Figure 1. Such soft neutrino spectrum, together with gamma-ray observations from {\it Fermi} allows for only a small fraction of its flux to be made by SFCRs with hard spectra $\alpha\la2.2$.

\begin{figure}\centering
\includegraphics[scale=0.5]{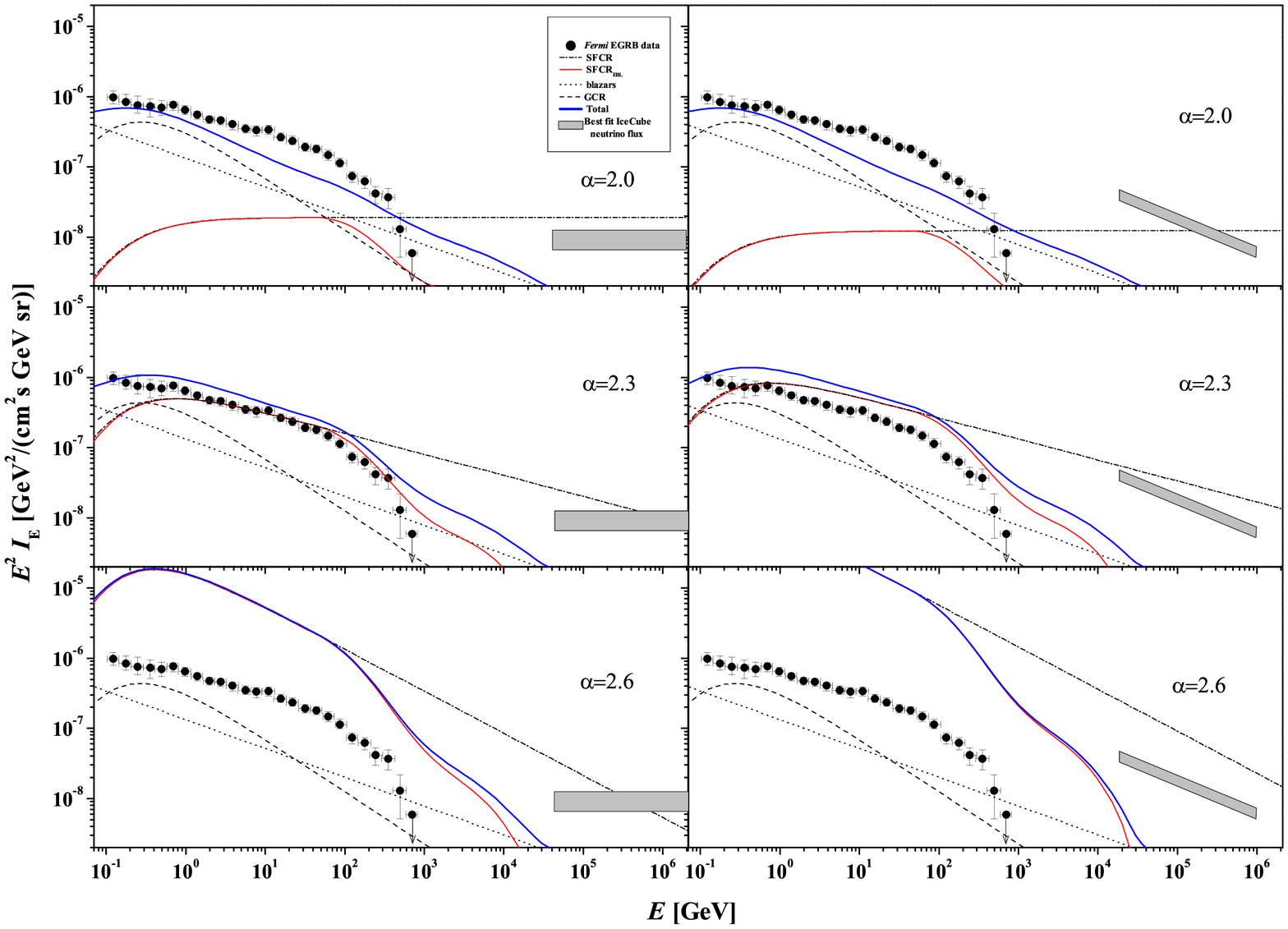}
\caption{SFCR contribution to the EGRB~\citep{AC14} calculated using models derived in~\cite{DP14}, with spectral indices $\alpha=2.0$ (top panels), $\alpha=2.3$ (middle panels), $\alpha=2.6$ (bottom panels). Our SFCR curves are all normalized to gamma-ray flux that corresponds to maximal neutrino flux allowed by the IceCube neutrino measurement for a given spectral index. Plots in the left column show normalization to the $\alpha_\nu=2$ IceCube spectrum~\citep{AAA2014}, and plots in the right column show SFCR curves normalized to latest $\alpha_\nu=2.46$ IceCube neutrino spectrum~\citep{AAA15}. Dashed dotted line represents SFCR gamma-ray emission without the EBL attenuation effects, while thin red solid line includes the EBL attenuation~\citep{GSP12}. We also plot blazar contribution (Abdo et al. 2010c; dotted line) and normal star-forming galaxies based on luminosity evolution model given in Fields et al. (2010; dashed line). Summarized contribution of these three components - SFCR with EBL attenuation, blazars, and normal galaxies is plotted using thick blue solid line. On our plot EBL attenuation is not included in blazar and normal galaxy spectra.}
\end{figure}

\cite{ZTG14} analyzed the contribution of galaxy clusters to the gamma-ray and neutrino background in a more model-dependent way and found that at best $10\%$ of the neutrino background can originate from SFCR interactions if all clusters are radio-loud with cosmic-ray spectrum $\alpha=2$, corresponding to gamma-ray flux at the level of only a few percent of the observed EGRB. Of course, given that our model is based on using detected neutrino flux as an upper limit to SFCR-made neutrinos, our results are consistent with ~\cite{ZTG14}. It should also be noted that
difference between the result of ~\cite{ZTG14} and the results of the work presented in this paper is also in part due to the difference between a typical cluster with respect to its SFCR contribution which is what our model gives, and the analysis based on the radio-loud clusters from~\cite{ZTG14} which do not necessarily have to be dominant sites for cosmic-ray acceleration at structure-formation shocks.
Compared to \cite{MAL13} who did not include source evolution and have found that clusters can contribute $\ga 30\%-40\%$ to the EGRB based on IceCube observation constraints, our results give lower flux limits.

Using the observed IceCube neutrino flux to constrain the SFCR contribution to the EGRB can further be used to constrain the gamma-ray emission of a typical (cosmic-average) cluster, such that clusters of that type dominate the SFCR gamma-ray background emission.
For example, starting from detected IceCube neutrino flux~\citep{AAA2014}, assuming that it originates from interactions of SFCRs with spectral index representative of strong shocks $\alpha=2$, we directly find the accompanying gamma-ray background emission. We can now estimate the expected flux of a typical cluster-type that dominates this background emission. For example, lets assume that this typical cluster type is similar in size and mass to galaxy cluster NGC1550, with total mass $M=0.68\times10^{14}\,M_\odot$, virial radius $r=0.77\,\mathrm{Mpc}$ and redshift $z=0.0123$~\citep{CH07}, which should be close to cosmic-average. Gamma-ray emission of this cluster is not expected to be dark mater dominated. Starting with neutrino background emission we can then estimate the gamma-ray flux of this cluster to be $F_{\gamma}=(0.2-3.6)\times10^{-12}\,\mathrm{phot}\,\mathrm{cm}^{-2}\mathrm{s}^{-1}$ in the \emph{Fermi} energy range $0.1-820\,\mathrm{GeV}$. For comparison, ~\cite{AC14b} calculated flux upper limits for $50$ galaxy clusters based on joint likelihood analysis of $4$ years of \emph{Fermi}-LAT data and using the assumption of the universal cosmic-ray model proposed by~\cite{PP10}. NGC1550 is one of the smallest clusters in their sample, and for this cluster they estimate the flux above $500\,\mathrm{MeV}$ to be $F_{\gamma}=4.9\times10^{-11}\,\mathrm{phot}\,\mathrm{cm}^{-2}\mathrm{s}^{-1}$. In the same energy range we derive NGC1550 flux of $F_{\gamma}=(0.09-1.4)\times10^{-12}\,\mathrm{phot}\,\mathrm{cm}^{-2}\mathrm{s}^{-1}$. On the other hand ~\cite{GDK14} give latest estimates for cluster flux upper limits in the energy range $0.8 - 100\,\mathrm{GeV}$ of $2.3\times10^{-11}\,\mathrm{phot}\,\mathrm{cm}^{-2}\mathrm{s}^{-1}$, which were based on stacking analysis of \emph{Fermi}-LAT photon count maps for the $78$ nearby rich clusters from the Two Micron All-Sky Survey ($2\mathrm{MASS})$ catalog. In this energy range we find the flux of our example cluster to be $F_{\gamma}=(0.6-9.9)\times10^{-13}\,\mathrm{phot}\,\mathrm{cm}^{-2}\mathrm{s}^{-1}$, which is at best over an order of magnitude below the estimate of ~\cite{GDK14}.

Similarly, we can estimate gamma-ray fluxes of other clusters, for example Coma cluster, but again with underlying assumption that this cluster is representative of a typical-type clusters that dominated the gamma-ray background. Thus, in the case of Coma,  using $z_\mathrm{Coma}=0.0232$, $M_\mathrm{Coma}=9.95\times10^{14}\,\mathrm{M}_\odot$, $r_\mathrm{Coma}=1.86\,\mathrm{Mpc}$~\citep{CH07} as values for redshift, total mass and virial radius respectively, we estimate the gamma-ray flux of this rich cluster to be around $F_{\gamma}=(0.09-1.5)\times10^{-11}\,\mathrm{phot}\,\mathrm{cm}^{-2}\mathrm{s}^{-1}$ in the $0.8 - 100\,\mathrm{GeV}$ energy range, which comes close (but not quite) to the~\cite{GDK14} upper limits. Though Coma is a rich cluster with mass about an order of magnitude larger than average cluster mass and thus it may not be a typical representative in terms of dominant SFCR gamma-ray emission source, it is an example of cluster where gamma-ray emission would be expected to be dominated by SFCR interactions rather than by dark matter, which makes it suitable when we want to compare its modeled emission with detection limits.

We can even further extend this analysis and constrain cosmic-ray acceleration efficiency. Using accretion shock models
like~\cite{PAFI06} we can estimate particle acceleration energy efficiency by comparing the
kinetic power of accreted gas that follows from the model, with the power that goes into SFCR particles which is constrained by the
detected neutrino and gamma-ray fluxes.  For example, if we assume that SFCRs
are dominantly accelerated in shocks that produce gamma-ray and neutrino spectra with index $\alpha=2.2$ \citep[corresponding
to Mach number $\sim 5 $][]{BE78,SCH10}, then from using the latest observed neutrino flux (with index $\alpha_\nu=2.46$)
as the upper limit it follows that the energy efficiency of accelerating cosmic rays  at these structures in the energy range
$1\,\mathrm{GeV}-1\,\mathrm{TeV}$ is  $\sim 40\%$, that is, we find that $\sim 40\%$ of energy that goes into shocks
gets converted into accelerated particles.  This is slightly higher efficiency compared to
results from numerical models of nonlinear diffusive shock acceleration of for example~\cite{KR13} who found that shocks of similar strength result in $\ga 10\%$
acceleration energy efficiency. If we on the other hand consider spectral index $\alpha=2$ to represent the SFCR spectra (corresponding
to Mach number $\sim 10 $) and use that to find the upper limit of SFCR flux as maximally allowed
by the observed extragalactic gamma-ray background and neutrino flux with index $\alpha_\nu=2$, we find that
in this case cosmic-ray acceleration energy efficiency is $< 1\%$. This is about an order of magnitude lower efficiency compared to
results from numerical model of \cite{KR13} who found that shocks of similar strength result in efficiency of $\sim 20\%$.
Such extreme changes in efficiency that follow from our analysis are of course due to the fact that our
constraint comes from the high-energy end that is fixed by observed neutrino fluxes, while most energy in
cosmic-rays comes from the low-energy end due to the power-law spectra.

\section{DISCUSSION AND CONCLUSION}
\label{sec:disc}

Gamma-ray observations have revealed the existence of an extragalactic gamma-ray background -- EGRB, which is still not completely explained. Formation of large-scale structures results in accretion shocks that can accelerate cosmic rays which should in turn leave an imprint on the EGRB. Interactions of these cosmic rays also produce charged pions which then decay and produce neutrinos. This common source directly links neutrino and gamma-ray astrophysics. Here we use the analytical models of gamma-ray emission of cosmic rays accelerated in accretion shocks around virialized structures over a range of redshifts derived in~\cite{DP14}, and normalize them to match the gamma-ray flux that would correspond to the observed high-energy IceCube neutrino flux~\citep{AAA2014b,AAA15} if they are assumed to come from SFCR interactions. This places an upper most limit on the SFCRs gamma-ray contribution to the EGRB observed by the \emph{Fermi}: $\la18\%$ for neutrino flux reported by \cite{AAA2014b}, that is $\la 12\%$ for neutrino flux reported by \cite{AAA15}, assuming SFCRs were accelerated in strong shocks ($\alpha=2.0$). This is the strongest yet model independent observational limit to this still unobserved cosmological cosmic-ray population.
Moreover, a solid limit to SFCR gamma-ray emission background in turn constrains the emission of typical objects that are the dominant source of SFCRs. The gamma-ray flux of the typical cluster estimated from our obtained limit was found to be around an order of magnitude lower than expected cluster emission from~\cite{GDK14}. We note however that this estimate is sensitive to the assumed spectral index of SFCRs, which, if adopted to match the high-energy IceCube neutrino flux of \cite{AAA2014b}, should be characteristic of strong shocks i.e. $\alpha=2$. For a softer spectra up to $\alpha \sim 2.3$ gamma-ray fluxes expected from SFCR interactions would be even higher, however in that case the high-energy neutrino flux would not be completely explained by SFCR population but would need at least one more source. Furthermore, if a typical cluster has mass and radius like the Coma cluster, radio observations of Coma cluster constrain its cosmic-ray spectral index to be closer to $\alpha=2.6$~\citep{BBR12} and the limits that can be derived from IceCube neutrino flux for such weak shocks become less constraining since the derived gamma-ray flux overshoots the EGRB data measured by the \emph{Fermi}-LAT. One of the major problems is actually identifying which type of clusters dominate accretion shock processing and are dominant sources of SFCR radiation, because no clear detection of clusters has been made yet.
However, using now new IceCube point-source limits from four years of data \citep{AAA2014b} together with known \emph{Fermi}-LAT \citep{AC14b} cluster limits, and observations of radio halos might prove to be sufficient to find strong predictions for SFCR emission of some clusters and that is the topic of our upcoming analysis.

\acknowledgments
We are thankful to John Beacom for valuable discussion. We are grateful to the anonymous Referee for valuable comments which helped improve this paper. The work of AD is supported by the Ministry of Science of the Republic of Serbia under project number 176005 and the work of TP is supported in
part by the Ministry of Science of the Republic of Serbia under project numbers 171002 and 176005.

\end{document}